\documentclass[prl,aps,floatfix,twocolumn,longbibliography]{revtex4-1}
\usepackage{epsfig}
\usepackage{amsmath}
\usepackage{amsfonts}
\usepackage{amssymb}
\usepackage{xcolor}
\usepackage[caption=false]{subfig}
\usepackage{graphicx}

\usepackage[colorlinks,linkcolor=blue,citecolor=green]{hyperref}
\usepackage[T1]{fontenc}
\usepackage{bm}

\begin{document}

\title{
Competing magnetic interactions in spin-1/2 square lattice: hidden order in Sr$_2$VO$_4$}
\author{Bongjae Kim$^{1}$}
\author{Sergii Khmelevskyi$^{1,2}$}
\author{Peter Mohn$^{2}$}
\author{Cesare Franchini$^{1}$}

\affiliation{$^{1}$ University of Vienna, Faculty of Physics and Center for Computational
Materials Science, Vienna A-1090, Austria}
\affiliation{$^{2}$ Center for Computational Materials Science, Institute for Applied Physics, Vienna University of
Technology, Wiedner Hauptstrasse $8$ - $10$, $1040$ Vienna, Austria}

\date[Dated: ]{\today}

\begin{abstract}
With decreasing temperature Sr$_2$VO$_4$ undergoes two structural phase transitions, tetragonal-to-orthorhombic-to-tetragonal,
without long-range magnetic order. Recent experiments suggest,
that only at very low temperature Sr$_{2}$VO$_{4}$ might enter some, yet unknown,
phase with long-range magnetic order, but without orthorhombic distortion.
By combining relativistic density functional theory with an extended spin-1/2 compass-Heisenberg model
we find an antiferromagnetic
single-stripe ground state with highly competing exchange interactions, involving a non negligible
inter-layer coupling, which places the system at the crossover between between the XY and Heisenberg picture.
Most strikingly, we find a strong two-site "spin-compass" exchange anisotropy
which is relieved by the orthorhombic distortion induced by the spin stripe order.
Based on these results we discuss the origin of the hidden order phase and the possible formation of a spin-liquid at low temperatures.
\end{abstract}

\maketitle

\preprint{}
%%%%%%%%%%%%%%%%%%%%%%%%%%%%%%%%%%%%%%%%%%%%%%%%%%%%%%%%%%

%%%%%%%%%%%%%%%%%%%%%%%%%%%%%%%%%%%%%%%%%%%%%%%%%%%%%%%%%
The Heisenberg model on a square lattice is one of the most
widely studied models in statistical physics. The applicability of this
model in modern solid-state physics has become popular after
the discovery of layered high-$T_c$ superconductors, since the parent
magnetically ordered compounds are often considered as quasi
two-dimensional (q2D) systems~\cite{Johnston1997},
and, more recently, for the interest in the interplay between magnetism and superconductivity
in Fe-based superconductors~\cite{Dai2012}.
In the 2D Heisenberg model the relative strength and competition of
antiferromagnetic nearest and next nearest neighbor interactions ($J_{1}$
and $J_{2}$) provides useful insights on the stability of
specific types of magnetic order: $J_{1}$ favors the N\'{e}el order (e.g., cuprates~\cite{Wan2009}),
$J_{2}$ favors the stripe (ST) order (e.g. few types of vanadates~\cite{Melzi2000,Melzi2001,Rosner2002,Nath2008}),
whereas a spin-liquid state is expected to emerge near the classical phase-boundary between the N\'{e}el and the ST orderings~\cite{Capriotti2001,Jiang2012,Li2012}.
The inclusion of additional interactions ($J_{3}$, $J_{4}$, \ldots) leads to more exotic states~\cite{Sindzingre2009},
for example the enigmatic nematic phases recently found in Fe-based superconductors which arises from the strong
competition between the ST and N\'{e}el order~\cite{Wang2016,Ma2009,Hu2012,Glasbrenner2015}.

Tetragonal Sr$_{2}$VO$_{4}$, isostructural to the high-$T_c$ parent compound La$_2$CuO$_4$ and similar to layered
vanadate Li$_{2}$VOSiO$_{4}$~\cite{Melzi2000}, provides an opportunity to explore the role of the various types
of magnetic interaction at play in a square lattice, owing to the presumably weak inter-plane interaction between spin-1/2 V$^{4+}$ ($d^{1}$) layers.
The complicated structural, magnetic, and electronic transitions observed in Sr$_{2}$VO$_{4}$, in fact, suggests a
competition and/or coexistence of different magnetic
phases~\cite{Zhou2007, Imai2005,Imai2006, Jackeli2009,Cyrot1990, Eremin2011,Yamauchi2015,Teyssier2016,Sugiyama2014}.
Upon cooling, the crystal structure evolves from tetragonal to an intermediate phase (at $T_{c2}\sim 140K$), and again to
a tetragonal phase ($T_{c1}\sim 100K$) with a larger $c/a$ ratio~\cite{Zhou2007}.
In the intermediate regime, an anomaly in the susceptibility is observed at $T_{M}\sim
105K$, which was initially thought to originate from magnetic order~\cite{Zhou2007}.
No signals of spin-orbital order was detected down to low temperature but, rather,
a possible
existence of a nematic phase, originating from the
competition between magnetic interaction and  Jahn-Teller (JT)  effect~\cite{Teyssier2016}.

\begin{figure}[t]
\begin{center}
\includegraphics[width=0.99\columnwidth,clip=true]{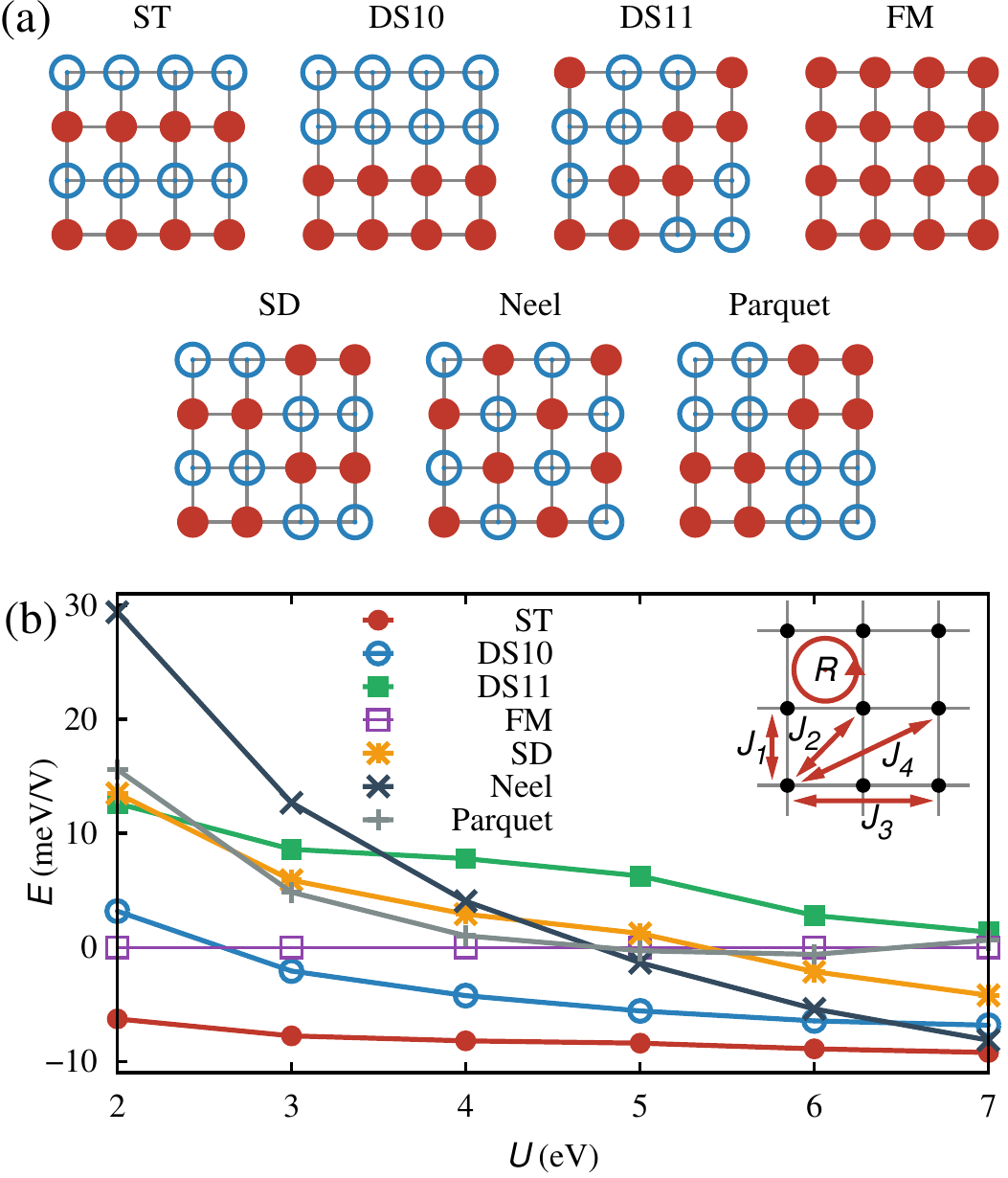}
\end{center}
\caption{ (a) Sketch of the different magnetic orderings used to fit the Heisenberg spin
Hamiltonian. Single stripe (ST), double stripe along [100] and [110]
(DS10 and DS11), ferromagnetic (FM), staggered dimer (SD), N\'{e}el,
and Parquet order. (b) Total energy as a function of $U$ within DFT+$U$ calculations.
Inset: schematic description of the exchange interactions.
}
\label{fig1}
\end{figure}

Several proposals have been put forwarded for the elusive low-T ground state:
orbitally ordered phase due to JT distortion~\cite{Zhou2007}, a competition between an
orbitally ordered parquet-type and a double-stripe (DS) magnetic ordering~\cite{Imai2005,Imai2006}
(DS10 in Fig.~\ref{fig1}(a)), an octupolar order driven by spin-orbit coupling (SOC)~\cite{Jackeli2009},
and  a N\'{e}el order with muted magnetic moment,
where the spin and orbital moments cancel each other~\cite{Eremin2011}.

The magnetic properties are highly dependent on the sample quality~\cite{Cyrot1990}. Recent experiments performed
with high quality samples have shed some light on the physics of Sr$_{2}$VO$_{4}$.
X-ray diffraction (XRD) experiments has revealed that the intermediate structure ($T_{c1}<T<T_{c2}$) is
orthorhombic phase~\cite{Yamauchi2015,Teyssier2016}.
Based on a muon spin rotation and relaxation ($\mu ^{+}$SR) study Sugiyama \emph{et. al} found that the actual magnetic ordering
temperature ($T_{N}\sim 10K$) is much lower than the temperature of the susceptibility
anomaly ($T_{M}\sim 105K$) and claimed that the role of the SOC
is marginal~\cite{Sugiyama2014}. This is consistent with the reported
persistence of inhomogeneous magnetic states down to 30 mK, where the sizeable
competition of ferromagnetic and antiferromagnetic correlations prevents the
system to develop a long range ordered phase~\cite{Yamauchi2015}.
The current understanding of the magnetic and structural properties of Sr$_{2}$VO$_{4}$
is thus still debated and a  commonly agreed explanation is still lacking.

In this Letter we investigate the peculiar magnetic behavior of Sr$_{2}$VO$_{4}$ by
first-principles calculations including relativistic effects. We show that
Sr$_{2}$VO$_{4}$  can be considered as a frustrated spin-1/2 Heisenberg
system with highly competing exchange interactions. Our data indicate that
the onset of the magnetic order is controlled by two factors with similar
order of magnitude: the relativistic two-site exchange anisotropy and a small, but
non-vanishing, inter-plane coupling. In particular, we argue that the absence
of a static magnetic order down to low-$T$ and the origin of the
tetragonal-orthorhombic transition, interpreted earlier either as nematic
transition or orbital order, can be understood as a competition of
anisotropic "spin-compass" exchange interactions, exchange magnetostriction and
antiferromagnetic interlayer coupling.

%-----------------------------------------------------------------------
%-----------------------------------------------------------------------

To study the magnetic interactions in Sr$_{2}$VO$_{4}$ we have performed first principles calculations
within the density functional theory (DFT) plus an on-site Hubbard $U$ using the Vienna Ab Initio Simulation Package~\cite{Kresse1993,Kresse1996}.
The experimental high-$T$ tetragonal structure~\cite{Rey1990} was modeled
with a supercell containing 16 unit cells, with which we have simulated different types of
spin orderings: single stripe (ST), double stripe along [100] and [110]
(DS10 and DS11, respectively), ferromagnetic (FM), staggered dimer (SD), N\'{e}el order,
and parquet order [see Fig.~\ref{fig1}(a)].
For the calculation of the exchange parameters we adopt the value $U$=5 eV, which is consistent with the
value obtained fully \emph{ab initio} within the constrained random phase approximation (cRPA).
Further methodological details are available in the Supplemental Materials~\cite{SM}.

Fig~\ref{fig1}(b) shows the DFT+U total energies for the considered spin orderings as a function of $U$.
The data for $U<$2~eV are omitted because for these values of $U$ DFT finds a metallic ground state, in disagreement with
experiment~\cite{Matsuno2003,Matsuno2005}.
We find that the ground
state is the ST configuration irrespective of the value of $U$. This is
surprising because in earlier literature the ST phase has never been considered
as a possible ground state magnetic structure, even though other layered oxides such as Li$_{2}$VOSiO$_{4}$
exhibit the ST order~\cite{Melzi2000}.
Our data shows that the relative energies does not depend strongly on $U$, and
with increasing $U$ the relative-energy window becomes narrower indicating that the
strength of the magnetic interactions are progressively weakened.
This can be expected since the  $Js$ are inversely
proportional to $U$, $J_{ij} \sim t_{ij}^{2}/U$, where $t_{ij}$ is the hopping between
two sites, $i$ and $j$. The shrinking of the energy window clearly implies
a strong competition among the various exchange interactions in action between neighboring spins
(see inset in Fig.~\ref{fig1}).

Independently on the magnetic state and on the $U$-value DFT deliver a spin moment on
V of about 1 $\mu_B$/V, which suggests a spin-1/2 local moment state, in agreement with the experimental
Curie-Weiss behavior of the high-$T$ magnetic susceptibility which provides an effective moment
close to 1.36~$\mu_{B}$~\cite{Sugiyama2014}.
However, Sr$_{2}$VO$_{4}$ does not show any magnetic order down to very low-$T$~\cite{Cyrot1990}.

The driving mechanism that stabilizes a magnetic
order at finite temperatures is one of the most important and subtle question related
to q2D magnetic compounds, as stated in the famous Mermin-Wagner theorem~\cite{Mermin1966}.
To analyze the physical reasons for the absence of magnetic order in Sr$_{2}$VO$_{4}$ we start by
mapping the different spin-ordered DFT total energies onto the following Heisenberg-like Hamiltonian,
\begin{widetext}
\begin{align}
H_{H}& =J_{1}\sum_{\alpha ,\langle ij\rangle }\mathbf{S}_{i\alpha }\cdot
\mathbf{S}_{j\alpha }+J_{2}\sum_{\alpha ,\langle \langle ij\rangle \rangle }%
\mathbf{S}_{i\alpha }\cdot \mathbf{S}_{j\alpha }+J_{3}\sum_{\alpha ,\langle
\langle \langle ij\rangle \rangle \rangle }\mathbf{S}_{i\alpha }\cdot
\mathbf{S}_{j\alpha }
+J_{4}\sum_{\alpha ,\langle \langle \langle \langle ij\rangle \rangle
\rangle \rangle }\mathbf{S}_{i\alpha }\cdot \mathbf{S}_{j\alpha }   \notag \\
& +J_{\bot}\sum_{\alpha ,\langle ij\rangle }\mathbf{S}_{i\alpha }\cdot \mathbf{S}%
_{j\alpha +1}+R\sum_{\alpha,~plaquette}[(\mathbf{S}_{i\alpha }\cdot \mathbf{S%
}_{j\alpha })(\mathbf{S}_{k\alpha }\cdot \mathbf{S}_{l\alpha })
+(\mathbf{S}_{i\alpha }\cdot \mathbf{S}_{l})(\mathbf{S}_{k\alpha }\cdot
\mathbf{S}_{j\alpha })+(\mathbf{S}_{i\alpha }\cdot \mathbf{S}_{k\alpha })(%
\mathbf{S}_{j\alpha }\cdot \mathbf{S}_{l\alpha })],
\label{hamiltonian}
\end{align}
\end{widetext}%
where $\langle \rangle $, $\langle \langle \rangle \rangle $, $\langle
\langle \langle \rangle \rangle \rangle $, and $\langle \langle \langle
\langle \rangle \rangle \rangle \rangle $ represent the sum over first, second,
third and fourth nearest neighbors (NNs), $R$ is the multisite ring exchange summed within a
square plaquette as shown in the inset in Fig.~\ref{fig1}(b), and the index $\alpha$ denotes
magnetic layers.
A similar procedure has been successfully applied for other 2D
square lattices such as the Fe-based superconductors~\cite{Glasbrenner2015},
Sr$_{2}$RuO$_{4}$~\cite{BKim2017}, and BaTi$_{2}$Sb$_{2}$O~\cite{Zhang2017}.

The estimated exchange interactions are listed in Table~\ref{table1}.
The stability of the ST configuration can be explained by the large AFM exchange coupling $J_{2}$=1.02 meV,
which is the dominating interaction~\cite{Shannon2006}, about twice larger than the first NN
FM-interaction $J_{1}$ (-0.65 meV).
However, the large values of $J_{3}$ and $J_{4}$ (-0.54 and 0.46, respectively) suggest that the
system deviates considerably from the $J_{1}-J_{2}$ model~\cite{moment}.
By extrapolating to large  $J_{2}$ the quantum Monte Carlo results obtained for the $J_{1}-J_{2}-J_{3}$
model~\cite{Mikheenkov2011}, one would expect that our data ($J_{2}/J_{1}=-1.57,$
$J_{3}/J_{1}=0.84$) should fall in the spin-liquid/spin-glass part of the
quantum spin-1/2 Heisenberg $J_{1}-J_{2}-J_{3}$ phase diagram.

Apart from the NNs Js, there are other two important terms in our spin Hamiltonian, the multisite ring interaction
$R$ and  the interlayer coupling $J_{_{\bot }}$.
Despite being small, the ring coupling $R$  is essential to explain the
energy difference between DS11 and the Parquet configurations (see Tab.\ref{table1}), because $R$
produces the same contribution to the energy for all orderings but DS11.
It would be therefore interesting to try to interpret the absence of a static classical
magnetic order in Sr$_{2}$VO$_{4}$ with a 2D quantum $J_{1}-J_{2}-J_{3}-J_{4}-R$ Heisenberg model.
Notably, the, ferromagnetic, interlayer coupling $J_{_{\bot }}$ is not negligible, -0.12 meV,
and brings the system on the 2D/3D threshold.
$J_{_{\bot }}$ favors the FM configuration and to a lesser extent DS10, DS11 and
Parquet configurations leaving the energy of ST and N\'{e}el order unchanged.

It has been argued that even small value of $J_{_{\bot }}$ can stabilize
the magnetic order with a rather high transition temperature~\cite{Jiang2012}.
However, due to the peculiarity of the body-centered tetragonal structure, where each spin-site is connected to
four spin-sites in the adjacent layer, it might happen that for a given in-plane
order the inter-plane interaction is fully frustrated. Examples of
such frustration are ordinary in-plane N\'{e}el order (e.g. cuprates)
or single stripe order (e.g. Fe-based pnictides). In Fe-based superconductors it has been
suggested that the critical role for the ordering might be played by a two-site magnetic
anisotropy~\cite{Ding1992} (note that for spin 1/2 systems
the single site anisotropy vanishes), which leads to the crossover between 2D
Heisenberg and either the 2D Ising or the XY-Kosterlitz-Thouless model behavior
(depending on sign of the anisotropy)~\cite{Johnston1997}.
For instance, recently it has been shown that the 3D magnetic order in the novel q2D compound Sr$_{2}$TcO$_{4}$
originates from a tiny dipole-dipole coupling that leads to an out-of-plane anisotropy associated with
a two-site exchange~\cite{Horvat2017}.
Considering that our data on Sr$_{2}$VO$_{4}$ shows that $J_{_{\bot }}$ leaves the ST order frustrated (i.e. unchanged)
in the following part of the paper we will consider  the possibility that the stabilization of a magnetic order at
finite temperature could originates from anisotropic effects. Specifically, we will
analyze the anisotropy of the exchange interactions due to SOC.

\begin{table}[b]
\caption{Total energies of the considered spin-ordered configurations  and
magnetic interactions of the ST ground state evaluated from the Heisenberg Hamiltonian (Eq.~\ref{hamiltonian}).
Data given in meV for $U$=5~eV.}
\label{table1}%
\begin {ruledtabular}
\begin{tabular}{cccccccc}
  ST   & DS10  &  N\'{e}el  & Parquet & FM & SD  & DS11 & \\
 \hline
 -8.39 & -5.56 &  -1.35     &  -0.28  & 0  & 1.21 & 6.26 & \\
 &      &       &      &       &       &       &     \\
 \hline
 $J_1$ & $J_2$ & $J_3$ & $J_4$ & $J_{\perp}$ & $R$ & $J^{xy}$  & $J^{zz}$ \\
\hline
 -0.65 & 1.02 & -0.54 & 0.46 & -0.12 & -0.28 & -0.71 & 1.21
\end{tabular}
\end {ruledtabular}
\end{table}

\begin{figure}[t]
\begin{center}
\includegraphics[width=0.99\columnwidth,clip=true]{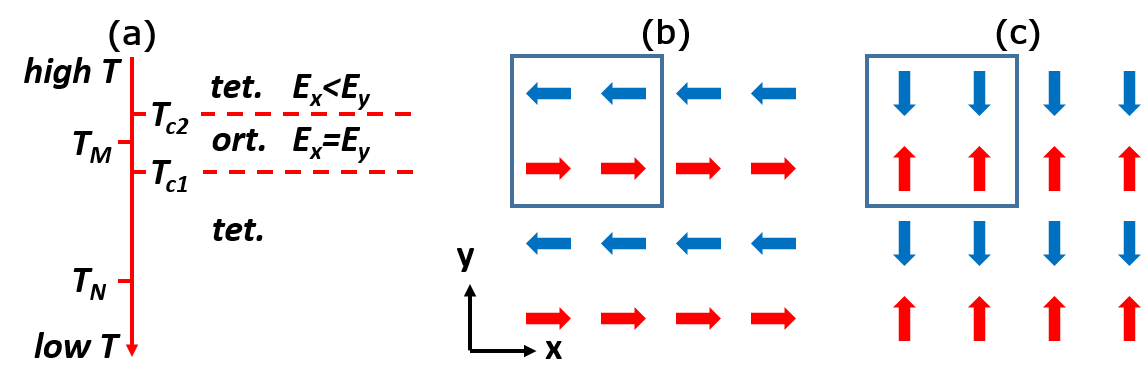}
\end{center}
\caption{ (a) Schematic diagram showing the transition temperatures and
structures. The system has a tetragonal structure with active anisotropic
exchange interactions for $T>T_{c2}$, while for the orthorhombic structure
for $T_{c2}>T>T_{c1}$ the anisotropy vanishes. Tetragonal symmetry is recovered
below $T_{c1}$. Stripe order is shown with moments along (b) (100) and (c)
(010), respectively. The squares denotes unit cell used. }
\label{fig2}
\end{figure}

To this aim, we have computed the total energy of the ST phase with the moments oriented
along the high-symmetry directions [001], [010] and [001] by means of magnetically
constrained DFT+U+SOC calculations~\cite{Liu2015} (see Fig.\ref{fig2}. We found that
$E_{x}$ is 2.75~meV and 2.63~meV (per V ion) lower than $E_{y}$ and $E_{z}$, respectively,
implying strong anisotropic effects.
As these differences cannot be explained by
single-site anisotropy (since we have a good spin-1/2 system) they must be
attributed to symmetric anisotropy exchange, which can be taken into account by including additional terms in
the spin Hamiltonian (Eq.~\ref{hamiltonian}). Within a first NN approximation the resulting spin Hamiltonian reads:

\begin{align}
H_{rH}&=H_{H}+\sum_{<nn>}J^{zz}S_{i}^{z}S_{j}^{z}  %\notag \\
&\pm\sum_{\substack{<nnx>\\<nny>}}J^{xy}(S_{i}^{x}S_{j}^{x}-S_{i}^{y}S_{j}^{y}),
\label{anisotropic}
\end{align}%
where $J^{zz}$ is the two-site Ising anisotropy (Kitaev term), and
the "spin-compass" exchange $J^{xy}$.
By mapping the total energies onto the above generalized spin Hamiltonian,
we obtain $J^{xy}$ = -0.71~meV and $J^{zz}$ =1.21~meV,
of the order of 10~K,  similar to the case of 4$d$ Sr$_{2}$RuO$_{4}$~\cite{BKim2017}.
The emergence of anisotropic terms demonstrates the
considerable role played by SOC even in 3$d$ oxides. However, we find
a negligible orbital moment ($<0.05$~$\mu _{B}$), which shows that SOC
contributes to the anisotropy of the system but does not form a spin-orbit
entangled phase~\cite{Eremin2011,Jackeli2009}.

The very strong Ising term $J^{zz}$
favors easy-plane anisotropy, putting the system close to the
Heisenberg/XY-model threshold since it does not lift the frustration of the ST
order. Thus Sr$_{2}$VO$_{4}$ is quite different from Sr$_{2}$TcO$_{4}$~\cite{Horvat2017}
and other similar q2D systems, where the magnetic order appears to be stabilized
by small negative $J^{zz}$ and a weak dipole-dipole interaction~\cite{Horvat2017}.
What is surprising and peculiar in Sr$_{2}$VO$_{4}$ is the existence of a strong spin-compass
anisotropy $J^{xy}$ which lifts the frustration of the ST phase and favors the
onset of magnetic order at finite temperatures.

From the calculated values of the dominant interactions we can now make some estimates and
provide a comparison with  experiment. At high temperatures the system exhibits a
Curie-Weiss (CW) behavior of the susceptibility with slightly different
CW temperatures $\theta _{CW}$ reported in literature: -24~K \cite{Sugiyama2014}, +24~K~\cite{Itoh1991},
and +47~K~\cite{Nozaki1991}. The mean-field value of $\theta _{CW}$ derived from our data (Tab.\ref{table1}) is
+10~K falling exactly in the middle of the experimental scattered data range.
If we estimate a mean-field ordering temperature of the ST phase we obtain +74~K,
which is somewhat below, but still close to the narrow temperature window,
100-140~K, covering the two structural ($T_{c2}$ and $T_{c1}$) transitions and
the magnetic susceptibility anomaly at ($T_{M}=105~K$)~\cite{Teyssier2016}.
Thus, the ST short-range order and
critical fluctuations are expected to develop exactly in the temperature
interval where the orthorhombic distortion exists.
But, there is a reason for Sr$_{2}$VO$_{4}$ to avoid a transition to ST order.

In q2D compounds with ST order, like Fe-pnictides and layered Li$_{2}$VOSiO$_{4}$~\cite{Melzi2000}, the
transition to the ST phase is associated with an orthorhombic
distortion, which is believed to break the frustration intrinsic to the ST phase~\cite{Han2009}.
The orthorhombic distortion in the Fe-pnictides~\cite{Huang2008}
appears at a temperature considerably higher than the magnetic phase transition implying that the short-range ST order is
responsible for the orthorhombic distortion even though long-range
magnetic order is absent.
In the following we show that the orthorhombic distortion in Sr$_{2}$VO$_{4}$
is a consequence of the magnetostriction effects in the ST phase.

To inspect the relation between magnetic and structural order we have fully relaxed
the high-$T$ tetragonal phase within DFT+U by imposing the ST order. As a result, we obtained a bifurcation
of the lattice parameters (a=3.971 \AA, b=3.960 \AA) and the formation of an orthorhombic phase, in agreement with XRD
measurements~\cite{Teyssier2016}. This transition is connected to the magnetostriction effect, namely
the coupling between non-relativistic isotropic exchange and crystal structure.
However, the inclusion of anisotropic effects via SOC leads to a very surprising result:
in the fully relaxed orthorhombic phase the anisotropic terms are
considerably reduced, $J^{xy}$ = 0.00~meV and $J^{zz}$ =0.38~meV, meaning that the
two-site compass term almost vanishes ($E_{x}=E_{y}$).
The orthorhombic distortion caused by magnetostriction in the ST phase
quenches the spin-compass anisotropy, which is the leading mechanism for the stabilization of
ST phase (lifting of the frustration) at finite temperature.
%\textbf{{BK: As the discussion is connected, it looks better not to divide the paragraph.}}

%-----------------------------------------------------------------------
%-----------------------------------------------------------------------

Based on our analysis we can now interpret the complicated transitions in Sr$_{2}$VO$_{4}$:
\emph{(i)} Starting well above the magnetic ordering temperature and decreasing the temperature
a strong short-range order, characteristic of a ST phase, develops within a tetragonal symmetry.
\emph{(ii)} This short-range ST order causes a tetragonal to orthorhombic structural transition and
simultaneously kills the spin-compass anisotropy; the system avoids a magnetic transition to the
ordered ST phase since the latter remains fully frustrated.
\emph{(iii)} When the temperature is lowered further, the system re-enters the tetragonal phase,
since other phases, like DS10, supported also by a non-frustrating inter-plane FM
coupling, become energetically equally favorable and partially destroy the short-range ST order.
\emph{(iv)} At this point the system enters an intermediate temperature range with different energetically competing
magnetic orders, in line with the complex behavior of the measured magnetic susceptibility;
\emph{(v)} Finally, at low temperature the system falls into some spin-liquid
or spin-glass states, where the FM phase becomes quasi-stable due to strong short-range FM interactions in the 1NN shells
both in-plane and inter-plane, in agreement with recent $\mu ^{+}$SR experiment~\cite{Yamauchi2015}.

In summary, we have shown that Sr$_{2}$VO$_{4}$ exhibits a peculiar magnetism, distinctively different
from other layered oxides. It displays a frustrated magnetic ground state  with significant
inter-plane interaction (placing it in the 2D/3D crossover regime), and strong
in-plane anisotropy (Heisenberg/XY model crossover), in particular the  "spin-compass" exchange term, which lifts the frustration
of the ST ground state. Anisotropic magnetic interactions, like the one observed in Sr$_{2}$VO$_{4}$,
are expected  be a pervasive phenomenon not only in 4$d$ and 5$d$, but also in 3$d$ compounds, in particular transition metal oxides,
and could play a key role in the onset of hidden magnetic orderings.
Our proposed scenario is compatible with experimental data and call for a full verification
by future experiments. For instance, we expect that nuclear magnetic resonance or inelastic neutron scattering  experiments on Sr$_{2}$VO$_{4}$ single crystals  could allow for a conclusive identification of the origin of the low-$T$ phases~\cite{Teyssier2016}
and should reveal the character of the magnetic short-range order effects in the three distinctive temperature intervals.

%%%%%%%%%%%%%%%%%%%%%%%%
%\bibliographystyle{apsrev4-1}
\bibliography{bibfile}

\begin{thebibliography}{43}
\bibitem [{mom()}]{moment}%
  \BibitemOpen
  \href@noop {} {\bibinfo  {journal} {Note that our calculated value of the
  moment fits well with spin-1/2 model with V$^{4+}$($d^1$) ion configuration:
  There are not much change in the moment upon varying $U$ (for ST order,
  0.96~$\mu_B$ for $U$=2~eV to 1.12~$\mu_B$ for $U$=7~eV), and for the same
  optical $U$=5~eV, the differences among diverse magnetic configurations are
  less than 0.02~$\mu_B$}\ }\BibitemShut {NoStop}%
\end{thebibliography}

\section{Acknowledgements}

We thank J. Sugiyama for fruitful discussions. This work was supported by
the Austrian Science Fund (FWF) projects ViCom (F4109-N28) and INDOX (11490-N19).
The computational results presented have been achieved using the Vienna Scientific Cluster (VSC).

%%%%%%%%%%%%%%%%%%%%%%%%%%%%%%%%%%%%%%%%%%%%%%%%%%%%%%%%%%%%%%%%%%
%%%%%%%%%%%%%%%%%%%%%%%%%%%%%%%%%%%%%%%%%%%%%%%%%%%%%%%%%%%%%%%%%%

%\vspace*{250px} \newpage

\end{document}